\documentclass[11pt]{article}
\usepackage{latexsym}
\usepackage{amsmath}
\usepackage{amsfonts}
\usepackage{graphics}
\newtheorem{Def}{Definition}
\newtheorem{The}[Def]{Theorem}
\newtheorem{Pro}[Def]{Proposition}
\newtheorem{Lem}[Def]{Lemma}
\newtheorem{Cor}[Def]{Corollary}
\newtheorem{Rem}{Remark}
\newtheorem{Exa}{Example}

\newenvironment{proof}
{\noindent
{\it Proof.}\ }{\ \(\Box\) \medskip}
\def\N{\mathbb{N}}
\def\Z{\mathbb{Z}}
\def\Q{\mathbb{Q}}
\def\R{\mathbb{R}}
\def\C{\mathbb{C}}
\def\r#1{{\rm r}_{#1}}
\def\val#1{{\rm val}_{#1}}
\begin{document}
\title{Numeration systems on a regular language : 
Arithmetic operations, Recognizability and Formal power series}
\date{January 17, 2000}
\author{Michel Rigo\\
Institut de Math\'{e}matiques, Universit\'{e} de Li\`{e}ge, \\
Grande Traverse 12 (B 37), B-4000 Li\`{e}ge, Belgium.\\
{\tt M.Rigo@ulg.ac.be}}
\maketitle
\begin{abstract}
Generalizations of numeration systems in which \(\N\) is 
recognizable by a finite automaton are obtained
by describing a lexicographically ordered infinite regular language 
\(L\subset \Sigma^*\). 
For these systems, we obtain a characterization of recognizable 
sets of integers in terms of rational formal series. We also show 
that, if the complexity of \(L\) is \(\Theta (n^l)\) (resp. 
if \(L\) is the complement of a polynomial language), then
 multiplication by \(\lambda\in \N\) preserves recognizability only if 
\(\lambda=\beta^{l+1}\) (resp. if \(\lambda\neq (\#\Sigma)^\beta\)) 
for some \(\beta\in \N\). 
Finally, we obtain sufficient conditions for the notions  
of recognizability and \(U\)-recognizability to be equivalent, where \(U\) is some 
positional numeration system related to a sequence of integers.
\end{abstract}

\section{Introduction}
According to \cite{LR}, a {\it numeration  system} 
is a triple \(S=(L,\Sigma,<)\) where \(L\) is an infinite regular 
language over a totally ordered alphabet \((\Sigma,<)\).
The lexicographic ordering of \(L\) gives a one-to-one correspondence 
\(r_{S}\) between the set \(\N\) of natural numbers and the language 
\(L\). A subset \(X\subset \N\) is called {\it \(S\)-recognizable} if 
\(\r{S}(X)\) is a regular subset of \(L\).

\medskip
We first characterize the \(S\)-recognizable subsets of \(\N\) 
in terms of rational series in the noncommuting variables \(\sigma 
\in \Sigma\) and with coefficients in \(\N\). In particular, we show 
that \(\sum_{n\in \N} n\, \r{S}(n)\) is rational (this kind of result 
is also discussed in \cite{BBG,CG}). Using classical 
results about rational series, we obtain a 
generalization of the fact given in \cite{LR} 
that ultimately periodic sets are \(S\)-recognizable 
for any numeration system \(S\).

\medskip
Our main purpose is related to the stability of the 
\(S\)-recognizability under arithmetic operations like addition and 
multiplication by a constant. If addition preserves the 
\(S\)-recognizability then multiplication by \(2\) 
 also preserves the \(S\)-recognizability. So, a natural question 
about the stability of the recognizability arises.
When does the multiplication by an integer \(\lambda\)
preserve the recognizability ?

It is well known that for positional numeration systems in base \(p\) 
the problem of addition and multiplication by a constant is completely 
settled. The \(p\)-recognizable sets are exactly those defined in the first 
order structure \(\langle \N,+,V_p\rangle\) (see for instance 
\cite{BH,BHMV}). 
It is obvious that addition and multiplication by a constant 
are definable in the Presburger arithmetic. Therefore, 
 \(p\)-recognizability is preserved. 

On the other hand, using the specific structure of the language \(a^* b^*\), 
it is shown in \cite{LR} that 
for the numeration system \(S=(a^*b^*,\{a,b\} ,a<b)\), the multiplication by 
a non-negative integer \(\lambda\) transforms the \(S\)-recognizable sets into \(S\)-recognizable sets 
if and only if \(\lambda\) is a perfect square. Then the multiplication 
by \(2\) does not preserve \(S\)-recognizability.

Notice that the language \(a^* b^*\) has a 
polynomial complexity (the complexity function \(\rho_L (n)\) of a language 
\(L\) counts the number of words of length \(n\) in \(L\)). 
So, it is natural to check whether a 
numeration system on a polynomial 
 language preserves the recognizability of a set after 
multiplication by a constant. For \(a^* b^*\), perfect squares play a 
special role. Does there exist a similar set for an arbitrary language in 
\(\Theta (n^l)\) ? 
We 
get the following result: if \(S\) is a numeration system built on 
a regular language with complexity in \(\Theta (n^l)\) then the 
multiplication by \(\lambda\) preserves the recognizability only if 
\(\lambda=\beta^{l+1}\) for some integer \(\beta\). 
As a consequence, the addition 
cannot be a regular map for numeration systems on polynomial regular 
languages.

In order to prove this, we proceed in two steps. 
In section \ref{NS3}, we assume that the complexity of the  
language is a polynomial of degree \(l\) with rational 
coefficients.  With such a language, we exhibit 
a subset \(X\) which is recognizable and we prove that \(\lambda X\) 
is not recognizable for any \(\lambda \in \N\setminus 
\{n^{l+1}:n\in \N\}\). In section \ref{NS4}, we consider the general 
case.

In this study of polynomial regular languages, we have obtained a 
interesting result about a special sequence associated to a language.
We denote by \(v_L (n)\), or simply \(v_n\) if 
the context is clear, the number of 
words of length not exceeding \(n\) belonging to \(L\).
In section \ref{NS4}, we show that if the complexity of \(L\) is 
 \(\Theta (n^l)\), then the sequence \((v_n/n^{l+1})_{n\in \N}\) 
converges to a strictly positive limit. It is  
surprising to notice that, in contrast, the sequence 
\((\rho_L (n)/n^l)_{n\in \N}\) generally does not converge.

The end of this paper is mainly related to exponential languages. 
In section \ref{NS5}, we consider numeration systems on the complement 
of a polynomial language. As in the polynomial case, we find a 
recognizable set \(X\) and constants \(\lambda\) such that \(\lambda 
X\) is not recognizable. Here, the \(\lambda\)'s are powers of the 
cardinality of the alphabet.

In the last section, we study relations between some positional 
numeration system \(U\) and a system \(S\) on a regular language 
\(L\). We give sufficient conditions for the equivalence of 
\(S\)-recognizability and \(U\)-recognizability. These conditions are 
strongly dependent on the language \(L\) and the recognizability of 
the normalization in \(U\). Using these conditions, we give two examples of 
numeration systems on an exponential language such that addition and 
multiplication by a constant preserve \(S\)-recognizability.
\section{Basic definitions and notations}

We denote by \(\Sigma^*\) the free monoid (with identity \(\varepsilon\)) 
generated by \(\Sigma\). For a set \(S\), \(\# S\) 
is the cardinality of \(S\) and for a string 
\(w\in \Sigma^*\), \(|w|\) is the length of \(w\).

Let \(L\subseteq \Sigma^*\) be a regular language;  
the minimal automaton of \(L\) is a \(5\)-tuple 
\(M_L=(K,s,F,\Sigma,\delta)\) where \(K\) is
the set of states, \(s\) is the initial state, 
\(F\) is the set  of final states and \(\delta:K\times
\Sigma \to K\) is the transition function. 
We often write \(k.\sigma\) instead of \(\delta(k,\sigma)\).  
Recall that the elements of \(K\) are the {\it derivatives} 
\cite[III.5]{E}
\[{w^{-1}.L=\{ v \in \Sigma^* \, : \, wv \in L \}},\, w\in\Sigma^*.\] 
The state \(k\) is equal to \(w^{-1}.L\) if and only if \(k=s.w\); 
\(w^{-1}.L\) being then the set \(L_k\) of words accepted by \(M_L\) from \(k\). In
particular, \(L=L_s\).

We denote \(u_l(k)\) the number \( \# (L_k \cap \Sigma^l)\) of words of length \(l\) belonging to \(L_k\)
and \(v_l(k)\) the number of words of length at most \(l\) belonging to \(L_k\), 
\[v_l(k) = \sum_{i=0}^l u_i(k).\]
Notice that the notations \(L_k\), \(u_l(k)\) and \(v_l(k)\) are relevant to any 
DFA (deterministic finite automaton) accepting \(L\).

The {\it lexicographic ordering} can be used to compare words of 
different length. Let \(x\) and \(y\) be two words. We say that 
\(x<y\) if \(\vert x\vert <\vert y\vert\) or if \(\vert x\vert =\vert 
y\vert\) and there exist letters \(\alpha<\beta\) such that 
\(x=w\alpha x'\) and \(y=w\beta y'\).

An extension of numeration systems in which the set of 
representations is regular is the following.
\begin{Def}{\rm A {\it numeration system} 
 is a triple \((L,\Sigma,<)\) where \(L\) is an infinite regular 
language over a totally ordered finite alphabet \((\Sigma,<)\) (see 
\cite{LR}).
The lexicographic ordering of \(L\) gives a one-to-one correspondence 
\(\r{S}\) between the set \(\N\) of natural numbers and the language \(L\).

For each \(n\in\N\), \(\r{S}(n)\) is the \((n+1)^{th}\)  word of \(L\)  with respect to the
lexicographic ordering and is called the {\it \(S\)-representation} of \(n\).
For \(w\in L\), we set \(\val{S} (w)=\r{S}^{-1} (w) \) and  
we call it the {\it numerical value} of \(w\).

The mappings \(\val{S}\) and \(\r{S}\) are sometimes called {\it 
ranking} and {\it unranking} in the literature.
}\end{Def}

This way of representing integers generalizes linear 
numeration systems in which \(\N\) is recognizable by finite automata. 
Examples of such systems are the numeration systems defined by a 
recurrence relation whose characteristic polynomial is the minimum polynomial 
of a Pisot number (i.e. an algebraic integer 
\(\alpha >1\) such that its Galois conjugates have modulus less than 
one) \cite{BH}. 
(Indeed, with this hypothesis, the set of representations 
of the integers is a regular language.) The standard numeration systems with 
integer base and also the Fibonacci system belong to this class.

\begin{Def}{\rm 
Let \(S\) be a numeration system. A subset \(X\) of \(\N\) is {\it \(S\)-recognizable} if \(\r{S} (X)\) 
is recognizable by a finite automaton.
}\end{Def}

Let \(S=(L,\Sigma,<)\) be a numeration system. Each \(k\in K\) for which \(L_k\) 
is infinite leads to the numeration system \(S_k=(L_k,\Sigma,<)\). 
The applications \(\r{S_k}\)  and \(\val{S_k}\) are simply denoted 
\(\r{k}\) and \(\val{k}\) if the context is clear. 
If \(L_k\) is  finite, the applications \(\r{k}\) and \(\val{k}\) 
are defined as in the infinite case but the domain of the
former restricts to \({\{0,\ldots ,\# L_k -1\}}\).

With these notations, we can recall a very useful proposition.
\begin{Lem}\label{lemme} {\rm \cite{LR}} 
Let \(S=(L,\Sigma,<)\)  and 
\(M=(K,s,F,\Sigma,\delta)\) be a DFA accepting \(L\). If \(\sigma w\) belongs 
to \(L_k\), \(k\in K\), \(\sigma \in \Sigma\), \(w \in \Sigma^+\), 
then
\[\val{k}(\sigma w)=\val{k.\sigma} (w)+v_{|w|} (k)-v_{|w|-1} (k.\sigma)
+\sum_{\sigma'<\sigma} u_{|w|} (k.\sigma').\ \Box \]
\end{Lem}

\section{Recognizable formal power series}\label{NS2}
Let \(R\) be a semiring, a formal power series \(T:\Sigma^*\to R\) 
can be written as a formal sum
\[T=\sum_{w\in \Sigma^*} (T,w)\, w.\]
We mainly adopt the terminology of \cite{BR} concerning semirings, 
rational and recognizable series. Recall that for each 
word \(u \in \Sigma^*\) and for each formal series \(T\), one 
associates the series \(u^{-1}T\) defined by
\[u^{-1}T=\sum_{w\in \Sigma^*} (T,uw)\, w.\]
In other words, \((u^{-1}T,w)=(T,uw)\).

\medskip
It is shown in \cite{BR} that the series \(\sum_{w\in X^*} \pi_2 (
w) \, w 
\in\N \langle\langle x\rangle\rangle \) is rational. In the last expression, \(X\) is the alphabet 
\(\{x_0,x_1\}\) and if \(w=x_{i_k}\cdots x_{i_0}\) then \(\pi_2(
w) =2^k \, i_k + \cdots + 2\, i_1+i_0\) is the numerical value in base 
two of \(w\). 

Here, we obtain the same result 
for any numeration system on a regular language. 
Another proof of this result can be found in \cite{CG} where 
complexity problems are discussed. 

\begin{Pro}\label{Rat} Let \(S=(L,\Sigma,<)\) be a numeration system. The 
formal series 
\[{\cal F}_S=\sum_{w \in L} \val{S} (w) \, w\in \N \langle\langle \Sigma 
\rangle\rangle\] is recognizable.
\end{Pro}
\begin{proof} Let \(M_L=(K,s,F,\Sigma,\delta)\) be the minimal 
automaton of \(L\). For \(k,l \in K\), \(\sigma \in \Sigma\), we 
introduce the following series of \(\N \langle\langle \Sigma 
\rangle\rangle\)
\[\begin{array}{ll}
T_k & = {\displaystyle \sum_{w \in L_k, w \neq \varepsilon}} [\val{k} (w)-v_{|w|-1} (k)] 
\, w \cr
U_{l,k} & = {\displaystyle \sum_{w \in L_l, w \neq \varepsilon}} u_{|w|} 
(k) \, w\cr
U_{l,k}' & = {\displaystyle \sum_{w \in L_l}} u_{|w|} (k) \, w\cr
V_{l,k} & = {\displaystyle \sum_{w \in L_l, w \neq \varepsilon}} v_{|w|-1} 
(k) \, w\cr
\end{array}\]
\[W_{k,\sigma}  = \left\{ \begin{array}{ll}
 [\val{k} (\sigma)-v_{0} (k)]\, \varepsilon & {\rm if}\ \sigma \in L_k \cr
 0 & {\rm otherwise.} \cr
\end{array}\right.
\]
If \(k,l \in K\), \(\alpha,\sigma \in \Sigma\), then we have the 
following relations
\[i)\ \sigma^{-1} T_k = T_{k.\sigma} + {\displaystyle 
\sum_{\sigma'<\sigma}} U_{k.\sigma,k.\sigma'} + W_{k,\sigma}\]
\[\begin{array}{rlrl}
ii)\ & \sigma^{-1} U_{l,k}= {\displaystyle 
\sum_{\alpha \in \Sigma}} U_{l.\sigma,k.\alpha}' & \ iii) &
\sigma^{-1} U_{l,k}'= {\displaystyle 
\sum_{\alpha \in \Sigma}} U_{l.\sigma,k.\alpha}'\\
iv)\ & \sigma^{-1} V_{l,k}= V_{l.\sigma,k}+U_{l.\sigma,k}' & 
\ v) & \sigma^{-1} W_{k,\alpha} = 0.\\
\end{array}\]

To check relation \(i)\), one has to compute \((T_k,\sigma w)\). 
Notice that \(\sigma w \in L_k\) iff \(w \in L_{k.\sigma}\).  
Use Lemma \ref{lemme} and treat the case \(w=\varepsilon\) separately.

For relations ii) and iii), if \(\sigma w\) belongs to \(L_l\) then \(w 
\in L_{l.\sigma}\) and 
\[(U_{l,k},\sigma w)=u_{|w|+1} (k) = \sum_{\alpha \in \Sigma} 
u_{|w|} (k.\alpha).\]

In \(iv)\), one observes that \(v_{|w|}(k)=v_{|w|-1}(k)+u_{|w|}(k)\). 
Relation \(v)\) is immediate.

Therefore the submodule \({\cal R}\) of \(\N \langle\langle \Sigma 
\rangle\rangle\) finitely generated by the series \(T_k\)'s, 
\(U_{l,k}\)'s, 
\(U_{l,k}'\)'s, \(V_{l,k}\)'s, \(W_{k,\sigma}\)'s is stable for the 
operation \(T\mapsto \sigma^{-1} T\), \(\sigma \in \Sigma\). By 
associativity of the operation \(T\mapsto w^{-1} T\), this module is 
stable. By \cite[Prop. 1, p. 18]{BR}, the series of \({\cal R}\) 
are recognizable.

To conclude the proof, notice that
\[T_k+V_{k,k}={\displaystyle \sum_{w \in L_k, w \neq \varepsilon}} 
\val{k}(w) \, w = \sum_{w \in L_k} \val{k}(w) \, w.\]
Indeed, if \(\varepsilon \in L_k\) then \(\val{k}(\varepsilon )=0\).
\end{proof}

\begin{Exa}{\rm We consider the numeration system 
\(S=(a^*b^*,\{a,b\},a<b)\). We obtain a linear representation 
\((\lambda,\mu,\gamma)\) for \({\cal F}_S\) : 
\[\lambda =\left(1\ 0\ 0\right),\ \mu (a)=\left(\begin{array}{ccc}
1&1&0\cr
0&1&1\cr
0&0&1\end{array}\right), \ \mu (b)=\left(\begin{array}{ccc}
1&1&1\cr
0&1&1\cr
0&0&1\end{array}\right),\ \gamma =\left(\begin{array}{c} 
0\cr1\cr1\end{array}\right)\]
where \(\mu:\{a,b\}^*\to \N^{3\times 3}\) is a morphism of monoids. 
Thus one has \[\val{S}(w)=\lambda\, \mu (w)\, \gamma.\]
}\end{Exa}

\medskip
Inspired by the definition of \(U\)-automata given in \cite{BH}, we 
have the following characterization of the regular subsets of a 
regular language.

\begin{Lem}\label{morph} Let \(L\subset \Sigma^*\) be a regular language 
and \(M_{L}=(Q_{L},s_{L},F_{L},\Sigma,\delta_{L})\) be its minimal automaton. 
If \(M_{K}=(Q_{K},s_{K},F_{K},\Sigma,\delta_{K})\) is the 
minimal automaton of a regular language \(K \subset L\) then there 
exists a morphism \(h\) of automata between \(M_{K}\) and  \(M_{L}\) defined as follows
\[h:Q_{K}\to Q_{L}, \]
\[
\left\{ \begin{array}{l}
h(\delta_{K} (q,\sigma))=\delta_{L} (h(q),\sigma),\ \sigma 
    \in \Sigma,\ q \in Q_{K}, \cr
h(s_{K})=s_{L}, \cr
h(F_{K})\subseteq F_{L}.\cr \end{array}\right.
\]
\end{Lem}
\begin{proof} A state of \(M_{K}\) is a derivative of \(K\) of the form 
\[u^{-1}.K=\{v\in \Sigma^* : uv \in K\}.\]
Since \(K\subset L\), then \(u^{-1}.K\subset u^{-1}.L\). We consider the 
morphism \(h:Q_{K}\to Q_{L}\) defined by \(h(q)=u^{-1}.L\) if 
\(q=u^{-1}.K\) for some \(u\). We can verify the properties of \(h\) 
using the definition of the minimal automaton \cite[III.5]{E},
\begin{enumerate}
\item \(\delta_{K}(q,\sigma)=\sigma^{-1}.q= 
\sigma^{-1}.(u^{-1}.K)\) for some \(u\in \Sigma^*\), \(q\in Q_{K}\), 
\(\sigma \in \Sigma\). 
So \(\delta_{K}(q,\sigma)=(u\sigma)^{-1}.K\) and \((u\sigma)^{-1}.L = 
\sigma^{-1}.h(q)=\delta_{L}(h(q),\sigma)\).
\item \(s_{K} =\varepsilon^{-1}.K\) and \(s_{L}=\varepsilon^{-1}.L\).
\item A state \(q=u^{-1}.K\) belongs to \(F_{K}\) if \(\varepsilon \in 
u^{-1}.K\) therefore \(\varepsilon \in u^{-1}.L\) and \(h(q)=u^{-1}.L \in 
F_{L}\).
\end{enumerate}
\end{proof}

With this lemma, we can generalize Proposition \ref{Rat} and obtain a 
characterization of the \(S\)-recognizable sets.

\begin{The} Let \(S=(L,\Sigma,<)\) be a numeration system, a set 
\(X \subseteq \N\) is \(S\)-recognizable if and only if the formal 
series 
\[\sum_{w\in \r{S}(X)} \val{S}(w) \, w \in \N \langle\langle \Sigma 
\rangle\rangle\]
is recognizable.
\end{The}
\begin{proof} The condition is sufficient. The support of a 
recognizable series belonging to \(\N \langle\langle \Sigma 
\rangle\rangle\) is a regular language \cite[Lemme 2, p. 49]{BR}. 

The condition is necessary. By Lemma \ref{morph}, one has a morphism 
\(h:M_X\to M_L\) where \(M_X\) (resp. \(M_L\)) is the minimal 
automaton of \(\r{S}(X)\) (resp. \(L\)). We proceed as in the proof 
of Proposition \ref{Rat}. Let \(K\) be the set of states of \(M_X\); 
for \(k,l \in K\), \(\sigma \in \Sigma\), we introduce the following 
series
\[\begin{array}{ll}
T_k & = {\displaystyle \sum_{w \in L_k, w \neq \varepsilon}} 
[\val{h(k)} (w)-v_{|w|-1} (h(k))] 
\, w \cr
U_{l,k} & = {\displaystyle \sum_{w \in L_l, w \neq \varepsilon}} u_{|w|} 
(h(k)) \, w\cr
U_{l,k}' & = {\displaystyle \sum_{w \in L_l}} u_{|w|} (h(k)) \, w\cr
V_{l,k} & = {\displaystyle \sum_{w \in L_l, w \neq \varepsilon}} v_{|w|-1} 
(h(k)) \, w\cr
\end{array}\]
\[W_{k,\sigma}  = \left\{ \begin{array}{ll}
 [\val{h(k)} (\sigma)-v_{0} (h(k))]\, \varepsilon & {\rm if}\ \sigma \in L_k \cr
 0 & {\rm otherwise.} \cr
\end{array}\right.
\]
We conclude as in Proposition \ref{Rat}.
\end{proof}

In \cite{LR}, it is shown that for any numeration system \(S\),  
arithmetic progressions are always \(S\)-recognizable. Using formal 
series, we can obtain a generalization of this result. Here,  the language 
\(L\) is not necessary lexicographically ordered. 
 
\begin{Pro}\label{pa} Let \(L\subset \Sigma^*\) be an infinite regular language and 
\(\alpha:L\to \N\) be a one-to-one correspondence. If 
\[T=\sum_{w\in L} \alpha(w) \, w\in \N \langle\langle \Sigma 
\rangle\rangle\]
is recognizable then \(\alpha^{-1}(p+\N\, q)\) is a regular language.
\end{Pro}
\begin{proof} Assume \(p=0\). Consider the congruence of the 
semiring \(\langle\N ,+,.,0,1\rangle\) defined by \(n\sim n+q\). We 
denote by \({\cal N}\) the finite semiring \(\N/\!\!\sim\) and by \(\varphi\) the 
canonical morphism \(\varphi:\N \to {\cal N}\). The characteristic series of 
\(L\), \( \underline{L} = \sum_{w\in L} w\), is recognizable (see 
\cite[Prop. 1, p. 51]{BR}). So 
\[ U=\varphi(T+\underline{L}) = \sum_{w\in L} \varphi (\alpha(w) +1)\, w 
\in {\cal N} \langle\langle \Sigma \rangle\rangle \]
is rational (see \cite[Lemme 1, p. 49]{BR}). Since \({\cal N}\) is finite and 
\(U\) is rational, the set
\[U^{-1}(\{\varphi (1)\}) = \{w \in \Sigma^* : (U,w)=\varphi (1)\} = 
\alpha^{-1}(\N \,q)\]
is a regular language (see \cite[Prop. 2, p. 52]{BR}). 

If \(p\neq 0\) and \(p<q\), then consider the series \(U=\varphi(T)\) 
and the set \(U^{-1}(\{\varphi (p)\})\).
\end{proof}

\begin{Cor} Arithmetic progressions are \(S\)-recognizable for any 
numeration system \(S\).
\end{Cor}
\begin{proof}
This is a direct consequence of Propositions \ref{Rat} and \ref{pa}.
\end{proof}

\begin{Rem}{\rm One can easily characterize the congruences \(\sim\) of the 
semiring \(\langle\N ,+,.,0,1\rangle\) with finite index \(q>1\). The 
canonical morphism is denoted by \(\varphi\).

First notice that \(\varphi (0)\neq \varphi (1)\). 
Since \(\N/\!\!\sim\) is finite, there exist \(x,y \in \N\) such that \(x+y \sim x\). 
Let
\[y_0=\min \{y>0\vert \exists x : x \sim x+y\} \ {\rm and}\ x_0 = \min 
\{ x | x\sim x+y_0\}.\]
For all \(n\in \N\) and \(i=0,\ldots , y_0-1\), one has \(x_0+i\sim 
x_0+i+n y_0\). It is obvious that if \(y_0>1\) then for \(i,j \in 
\{0,\ldots , y_0-1\}\), \(i\neq j\), one has \(x_0+i \not\sim x_0+j\).
By definition of \(x_0\) and \(y_0\), if \(z<x_0\) then \(\varphi^{-1} \varphi (z) =\{z\}\)

Therefore the congruences of \(\N\) with finite index are generated 
by the relation \(n\sim n+y_0\) for \(n\) sufficiently large. So we cannot 
refine Proposition \ref{pa} with the same kind of proof because it 
uses explicitely the finiteness of \(\N/\!\!\sim\).
}\end{Rem}

\section{Multiplication for exact polynomial languages}\label{NS3}

In \cite{LR}, we proved that
for the numeration system \(S=(a^*b^*,\{a,b\} ,a<b)\), the multiplication by 
a non-negative integer \(\lambda\) transforms the \(S\)-recognizable sets into \(S\)-recognizable sets 
if and only if \(\lambda\) is a perfect square. 

In this section, we study the family 
of regular languages with polynomial complexity function. 
This step contains the main ideas leading to the case of an arbitrary 
polynomial language (i.e. a language with complexity function bounded by 
a polynomial). But it is simpler to handle since we only 
deal with polynomials.
\medskip

\begin{Lem}\label{fctcr} Let \(f:\N\to \N\) be a strictly increasing function such 
that \(f(\N)\) is a finite union of arithmetic progressions (i.e. 
there exist \(y_0\) and \(\Gamma\) such that \(\forall y\ge y_0\), \(y\in 
f(\N) \Leftrightarrow y+\Gamma \in f(\N)\)). 

Let \(k=f^{-1}(y_0+\Gamma )-f^{-1}(y_0)\). For all \(x\ge 
f^{-1}(y_0)\), \(n\in \N\), 
\[f(x+nk)=f(x)+n\Gamma.\]    
\end{Lem}
\begin{proof} Let \(x_0=f^{-1} (y_0)\). We have by definition of \(k\),
\[f(x_0+k)=f(f^{-1}(y_0)+k)= f(f^{-1}(y_0+\Gamma))=f(x_0)+\Gamma.\] 
It is sufficient to show that if \(x\ge x_0\) then
\[f(x+k)=f(x)+\Gamma\Rightarrow f(x+k+1)=f(x+1)+\Gamma.\]
Since \(f\) is strictly increasing, \(f(x+k+1)>f(x+k)=f(x)+\Gamma\). 
Since the characteristic sequence of \(f(\N)\) is ultimately 
periodic, there exists \(v\ge x_0\) such that \(f(v)=f(x+k+1)-\Gamma>f(x)\). 
Then \(v\ge x+1\). There exists \(u\in \N\) such that 
\(f(u)=f(x+1)+\Gamma > f(x)+\Gamma = f(x+k)\).

Now, assume that \(v>x+1\). Therefore \(f(v)>f(x+1)\) and
\[f(x+k+1)=f(v)+\Gamma>f(x+1)+\Gamma=f(u)>f(x+k).\] 
So we have \(x+k+1>u>x+k\) which is a contradiction and \(v=x+1\). 
\end{proof}

\begin{Def}{\rm The {\it complexity 
function} of a language \(L\subseteq \Sigma^*\) is 
\[\rho_{L}: \N \to \N : n \mapsto \# (\Sigma^n \cap L).\]
In the following, we assume that we deal with ``true" complexity functions, 
i.e. if \(\rho_L\) is a polynomial belonging to \(\Q[x]\) and \(n\in \N\) 
then \(\rho_L (n)\) is a non-negative integer. We equally use 
the notation \(\rho_L (n)\), \(u_n(s)\) or even \(u_n\) provided the 
context is clear.
}\end{Def}

The next lemma will be useful when applied to a complexity function.

\begin{Lem}\label{pol} If \(H\) is a polynomial such that \(\forall n 
\in \N\setminus\{0\},\ H(n)\in \Z\) then \(H(\Z)\subseteq \Z\).
\end{Lem}
\begin{proof} We proceed by induction on the degree of \(H\). 
If \(H\) is a polynomial of degree one then one has 
\(H(n)=a\, n+b\) with \(a,b\in \Z\) and \(H(\Z)\subseteq \Z\). 

Assume that the result holds for polynomials of degree \(k\ge 1\). If 
\(H\) is a polynomial of degree \(k+1\), then there exists a polynomial 
\(R\) of degree \(k\) such that \(\forall n\ge 1\), 
\(R(n)=H(n+1)-H(n)\in\Z\). 
Therefore \(R(\Z)\subseteq \Z\) and \(H(0)=H(1)-R(0)\in \Z\). We can 
conclude by induction on \(n<0\) because \(H(n)=H(n+1)-R(n)\).
\end{proof}

\begin{The}\label{mul} Let \(L\subset \Sigma^*\) be a regular language such that
\[\rho_L (n)=\left\{\begin{array}{cl}
a_l \, n^l + \cdots + a_1 \, n + a_0 & \ {\rm if}\ n>0 \cr
1 & {\rm otherwise}\end{array}\right.\]
where the \(a_i\)'s belong to \(\Q\) and \(a_l>0\). 
Let \(\prec\) be an ordering of the alphabet \(\Sigma\) and \(S=(L,\Sigma,\prec)\) 
be the corresponding numeration system.

If \(\lambda \in \N\setminus \{n^{l+1}: n\in \N\}\), then there exists a subset \(X\) of 
\(\N\) such that \(\r{S}(X)\) is regular and that \(\r{S}(\lambda \, X)\) is not.
\end{The}
\begin{proof} One can build a polynomial
\(P \in \Q[x]\) of degree 
\(l+1\)  such that \(P(0)=0\) 
and for all \(n\ge 1\), \(P(n+1)=P(n)+\rho_L (n)\). 

Indeed, let \(P(x)=b_{l+1} \, x^{l+1}+\cdots + b_1\, x + b_0\). The 
conditions on 
\(P\) gives the following triangular system
\[
\left\{\begin{array}{lcl}
a_l & = & b_{l+1}\, (l+1) \cr
a_{l-1} & = & b_{l+1}\,(l+1)\, \frac{l}{2} + b_l\, l \cr
 & \vdots & \cr
a_0 & = & b_{l+1} + \cdots + b_1 \cr 
b_0 & = & 0. \cr \end{array}\right.
\]
This polynomial \(P\) has some useful properties.  
We have the polynomial identity \(P(x+1)=P(x)+\rho_L(x)\) for \(x\in 
\N\setminus\{0\}\). Then it holds for \(x\in \R\)  if we extend the 
definition of \(\rho_L\) to \(\rho_L:\R\to\R : x\mapsto a_l \, x^l + \cdots + a_0\). 
By Lemma \ref{pol}, \(P(1)=\rho_L(0)=a_0 \in \Z\).
One shows by induction on \(n\in \N\) that \(P(n)\) (resp. \(P(-n)\)) 
is an integer since \(\rho_L(\N)\subset \N\) (resp. since 
\(\rho_L(\Z)\subset \Z\) by Lemma \ref{pol}).

Let \(x\in \N\setminus \{0\}\), notice that
\begin{equation}\label{lg}
 |\r{S}(x)| = n \Leftrightarrow x \in [P(n)-a_0+1,P(n+1)-a_0].
\end{equation}
Indeed, an integer \(x\) has a representation of length \(n\) if 
\(v_{n-1} \le x < v_n\) and 
\[v_n =\sum_{i=1}^n \rho_L (i) + 1 = \sum_{i=1}^n [P(i+1)-P(i)]+1 = 
P(n+1)-P(1)+1.\]
Notice that \(\r{S}(P(\N))\) is a translation of the set \({\cal 
I}(L,<)\) of the first words of each length. Therefore \(X=P(\N)\) 
is \(S\)-recognizable, see \cite{LR,Sh}.

Let \(\lambda\in\N\setminus\{0,1\}\). Our aim is to show that 
\(\lambda \, P(\N)\) is not \(S\)-recognizable.

For \(n\) large enough, we first show that
\[n \le |\r{S}(\lambda \, P(n))|< \lambda^{1/l} n.\]
The first inequality is obvious. 
In view of (\ref{lg}), to satisfy the second inequality, one must check 
whether   
\[\lambda \, P(n) < P(\lambda^{1/l} \, n)-a_0+1.\]
We can write \(P(n)\) as \(b_{l+1}\, n^{l+1} + Q(n)\) with 
\(b_{l+1}>0\) and \(Q\) being  a polynomial of degree not exceeding 
\(l\). Then,
\[\begin{array}{ll}
  & P(\lambda^{1/l} \, n)-\lambda \, P(n)-a_0 +1\cr
= & b_{l+1} (\lambda^{1/l} \, n)^{l+1} - \lambda \, b_{l+1} \,
n^{l+1} + Q(\lambda^{1/l} \, n+1) - \lambda \, Q(n) - a_0 +1.\cr
\end{array}\]
The coefficient of \(n^{l+1}\) is \(b_{l+1}\, (\lambda^{(l+1)/l} - 
\lambda)>0\). So, there exists \(n_0\) such that for all \(n\ge 
n_0\), this polynomial expression of degree \(l+1\) is strictly 
positive and \(|\r{S}(\lambda \, P(n))|< \lambda^{1/l}\, n\).

If \(n\) is sufficiently large, we show that
\[|\r{S}(\lambda \, P(n+1))|>|\r{S}(\lambda \, P(n))|.\]
Let \(i=|\r{S}(\lambda \, P(n))|\). In view of (\ref{lg}), one 
has to verify that \[\lambda \, P(n+1) > P(i+1)-a_0.\]
By definition of \(P\) and by (\ref{lg}), one has 
\[\lambda \, P(n+1) = \lambda \, P(n) + \lambda \, 
\rho_L(n) > P(i) - a_0 + \lambda \, \rho_L(n).\] 
Therefore it is sufficient to check whether 
\(P(i) - a_0 + \lambda \, \rho_L(n) > P(i+1) - a_0\), 
which occurs if and only if
\[\lambda \, \rho_{L} (n) - \rho_L (i) = a_l \, (\lambda n^l - i^l) + 
\cdots + a_k \, (\lambda n^k - i^k) +\cdots + a_0\, (\lambda -1)>0.\]
To verify that this inequality holds, remember that \(a_l>0\) and for \(n\ge n_0\), \(1\le 
\frac{i}{n}< \lambda^{1/l} \). Thus one studies the quotient \(\frac{\lambda \, \rho_{L} (n) - \rho_L 
(i)}{n^l}\) when \(n\to +\infty\),
\[\underbrace{a_l \, \left[\lambda  - 
\left(\frac{i}{n}\right)^l \right]}_{>0} + \cdots + 
\underbrace{\frac{a_k}{n^{l-k}}}_{\to 0} \, \underbrace{\left[\lambda - 
\left(\frac{i}{n}\right)^k\right]}_{is\ bounded}  + \cdots + 
\underbrace{\frac{a_0}{n^l}}_{\to 0} \, (\lambda -1).\]
So there exists \(n_0'\ge n_0\) such that for all \(n\ge n_0'\), 
\(|\r{S}(\lambda \, P(n+1))|>|\r{S}(\lambda \, P(n))|\).

Assume that \(\r{S}(\lambda \, P(\N))\) is regular then the set 
\(|\r{S}(\lambda 
\, P(\N))|\) is a finite union of arithmetic progressions.
We may apply Lemma \ref{fctcr}; indeed, the function \(|\r{S}(\lambda 
\, P(.))|\) is strictly increasing in \(\{n : n\ge n_0'\}\) and there 
exist \(l_0\) and \(\Gamma_\lambda\) (simply written \(\Gamma \)) such 
that \(\forall l\ge l_0,\  l\in |\r{S}(\lambda 
\, P(\N))| \Leftrightarrow l+\Gamma \in |\r{S}(\lambda \, P(\N))|\). 
Let \(n_1 \ge n_0'\) be such that \(|\r{S}(\lambda \, P(n_1))|> l_0\). 
By Lemma \ref{fctcr}, there exists \(k_\lambda\) (simply written 
\(k\)) such that for all 
\(n\ge n_1\) and for all \(\alpha \in \N\), 
\[|\r{S}(\lambda \, P(n+\alpha k))|=|\r{S}(\lambda \, P(n))|+\alpha 
\Gamma.\]
Let \(i=|\r{S}(\lambda \, P(n))|\). In view of (\ref{lg}), one 
has 
\[P(i+\alpha \Gamma)-a_0 +1 \le \lambda\, P(n+\alpha k) \le P(i+\alpha \Gamma+1)-a_0.\]
Since \( \lambda\, P(n+\alpha k) - P(i+\alpha \Gamma)+a_0 -1 \) must be 
positive for all \(\alpha \in \N\), the coefficient of the greatest 
power of \(\alpha\), \(\alpha^{l+1}\), must be strictly positive. 
This coefficient is
\[\lambda \, b_{l+1} \, k^{l+1} - b_{l+1} \, \Gamma^{l+1}\]
and we have the condition 
\[k>\frac{\Gamma}{\lambda^{1/(l+1)}}.\]
Notice 
that the coefficient vanishes only if \(\lambda = \left( 
\frac{\Gamma}{k} \right)^{l+1}\). By hypothesis, this case is 
excluded (notice that \(\frac{\Gamma}{k}\in \Q\backslash \N \Rightarrow 
(\frac{\Gamma}{k})^{l+1} \not\in \N\)).

But \( \lambda\, P(n+\alpha k) - P(i+\alpha \Gamma+1)+a_0\) must be 
negative for all \(\alpha \in \N\). The coefficient of the greatest 
power of \(\alpha\) is also \(\lambda \, b_{l+1} \, k^{l+1} - b_{l+1} \, 
\Gamma^{l+1}\) and must be strictly negative. 
Then we have simultaneously the condition 
\[k<\frac{\Gamma}{\lambda^{1/(l+1)}},\] which leads to a contradiction.
\end{proof}

In Theorem \ref{mul}, we exhibit a recognizable set 
\(X=P(\N)\) such that \(|\r{S}(\lambda\, P(\N))|\) is not a finite 
union of arithmetic progressions. When we consider the case 
\(\lambda=\beta^{l+1}\), \(\beta \in \N\setminus \{0,1\}\), we cannot 
find easily a subset \(X\) which is recognizable and such that 
\(\lambda\, X\) is not. 

The next proposition shows that 
\(|\r{S}(\beta^{l+1} P(\N))|\) is a finite 
union of arithmetic progressions whether \(\rho_L\) is a polynomial of 
degree \(l\).

\begin{Pro} With the assumptions and notations of Theorem \ref{mul}, 
there exists \(C\in \Z\) such that for \(n\) large enough,
\[|\r{S}(\beta^{l+1} P(n))|=\beta\, n+C.\]
\end{Pro}
\begin{proof} In the proof of Theorem \ref{mul}, we introduced a 
polynomial \(P(x)=b_{l+1} \, x^{l+1}+\cdots + b_1\, x\) such 
that \(P(n+1)-P(n)=\rho_L (n)\). In view of 
(\ref{lg}), we have to find an integer \(C\) such that for \(n\) large 
enough
\begin{eqnarray}\label{t1}
P(\beta\, n+C+1)-a_0-\beta^{l+1} \, P(n)&\ge &0 \\
\label{t2}
\beta^{l+1} \, P(n)-P(\beta \, n+C)+a_0-1 &\ge &0.
\end{eqnarray}
The coefficient of \(n^{l+1}\) vanishes in (\ref{t1}) and 
(\ref{t2}). The coefficient of \(n^l\) in (\ref{t1}) is
\(\beta^l\, [a_l\, (C+1)+b_l\, (1-\beta)]\) 
with \(a_l=b_{l+1}\, (l+1)\). It is strictly increasing with \(C\) 
and equals zero for
\[C=C_1:=\frac{b_l\, (\beta -1)-a_l}{a_l}.\]
The same coefficient in (\ref{t2}) is 
\(-\beta^l\, [a_l\, C+b_l\, (1-\beta)]\). It is strictly decreasing  
with \(C\) and equals zero for \(C=C_2:=C_1+1\).

If \(C_1\) and \(C_2\) are not integers then there exists \(C\in 
]C_1,C_2[ \cap \Z\) such that the coefficients of terms of maximal degree 
are both strictly positive. 

Otherwise, one has to consider the integer case \(C=C_1\) or \(C=C_2\) 
(it is obvious that any other \(C\) leads to a strictly negative expression for 
(\ref{t1}) or (\ref{t2})). Moreover, if \(C=C_1\) (resp. \(C=C_2\))
then (\ref{t2}) (resp. (\ref{t1})) is satisfied for \(n\) large 
enough. 

Notice that for \(i=1,\ldots ,l-1\) the coefficient of 
\(n^i\) in (\ref{t1}) with \(C=C_1\) is the opposite of the coefficient of 
\(n^i\) in (\ref{t2}) with \(C=C_2\) since \(C_2=C_1+1\). Notice also 
that the independent term in (\ref{t1}) for \(C=C_1\) is 
\(P(C_2)-a_0\). In (\ref{t2}) for \(C=C_2\) this term is \(-P(C_2)+a_0-1\). 
Thus we can write (\ref{t1}) with \(C=C_1\) as
\[ A_{l-1}\, n^{l-1} + \cdots + A_1\, n + P(C_2)-a_0 \]
and (\ref{t2}) with \(C=C_2\) as
\[- A_{l-1}\, n^{l-1}- \cdots - A_1\, n -P(C_2)+a_0-1.\]

If there exists \(i\) such that \(A_i\neq0\) then let \(j=\max_{A_i\neq 0} 
i\). If \(A_j>0\) (resp. \(A_j<0\)) then one takes \(C=C_1\) (resp. \(C=C_2\)).

Now, assume that \(A_i=0\) for \(i=1,\ldots,l-1\). If \(P(C_2)-a_0\ge 
0\) then one takes \(C=C_1\). Otherwise, \(-P(C_2)+a_0\) is a striclty 
positive integer (remember the properties of \(P\) obtained in the proof 
of Theorem \ref{mul}). 
Therefore \(-P(C_2)+a_0-1\ge 0\) and one takes \(C=C_2\).
\end{proof}
\section{Multiplication and polynomial languages}\label{NS4}
Here we obtain the generalization of Theorem \ref{mul} 
for an arbitrary regular language of polynomial complexity. 
In the same time, we show that the sequence \((v_n/n^{l+1})_{n\in\N}\) converges 
if the complexity of \(L\) is \(\Theta(n^l)\).

Let us recall some notations. Let \(f(n)\) and \(g(n)\) be two functions, it 
is said that \(f(n)\) is \(O(g(n))\) if there exist positive constants 
\(c\) and \(n_0\) such that for all \(n\ge n_0\), \(f(n)\le c\, g(n)\); 
\(f(n)\) is \(\Omega (g(n))\) if there exists a strictly positive constant 
\(c\) and an infinite sequence \(n_0,n_1,\ldots ,n_i,\ldots \) such 
that for all \(i\in \N\), \(f(n_i)\ge c\, g(n_i)\). The function \(f(n)\) is 
 \(\Theta (g(n))\) if \(f(n)\) is \(O(g(n))\) and \(\Omega 
 (g(n))\). Let \(x\in A^*\) and \(y\in B^*\), with \(A\) and \(B\) 
 two finite alphabets. If \(\vert x\vert = \vert y \vert + i\), 
 \(i\in \N\) then \((x,y)^\#=(x,\#^i y)\) where \(\#\) is a new symbol 
 which does not belong to \(A\cup B\). If \(\vert y \vert=\vert 
 x\vert+i\) 
 then \((x,y)^\#=(\#^i x, y)\). This operation can be extended to 
 \(n\)-uples of words. Let \(R\) be a relation over \(A^*\times 
 B^*\). We say that \(R\) is {\it regular} if \(R^\#\) is a regular 
 language. This definition can be extended to \(n\)-ary relations. 
 A map is {\it regular} if its graph is regular.

\begin{The}\label{Grand} Let \(L\subset \Sigma^*\) be a regular language such that 
\(\rho_L(n)\) is \(\Theta (n^l)\) for some integer \(l\).  
If \(\lambda\in \N\setminus\{n^{l+1}:n\in\N\}\),  
then there exists a subset \(X\) of 
\(\N\) such that \(\r{S}(X)\) is regular and that 
\(\r{S}(\lambda \, X)\) is not.
\end{The}

This theorem has a direct corollary.

\begin{Cor} Under the assumptions of Theorem \ref{Grand}, the addition is 
not a regular map (i.e. the graph of the application \((x,y)\mapsto 
x+y\) is not regular).
\end{Cor}
\begin{proof} By Theorem \ref{mul}, there exists a subset \(X\) of 
\(\N\) such that \(X\) is \(S\)-recognizable and \(2X\) is not. 
Assume that the 
graph of the addition
\[\hat{\cal G}=\{(\r{S}(x),\r{S}(y),\r{S}(x+y))^{\#} : x,y \in \N\}\]
is regular. Let \(p_3\) be the canonical 
homomorphism defined by \(p_3(x,y,z)=z\). It is clear that the set 
\(A=\{(\r{S}(x),\r{S}(x),w)^\# : x \in X, w \in \Sigma^*\}\) is regular. 
Therefore
\[A\cap \hat{\cal G} = \{ (\r{S}(x),\r{S}(x),\r{S}(2x))^\# : x\in X\}\] 
is regular.  
Thus \(p_3(A\cap \hat{\cal G})=\r{S}(2X)\) is also regular, a contradiction. 
\end{proof}

In the following, we will use the term of \(k\)-tiered word 
and the results obtained in \cite{Yu} about the complexity of regular polynomial 
languages.

\medskip
The first lemma is just a refinement of \cite[Lemma 1]{Yu}. We simply 
remark that one can consider an ultimately periodic sequence \(n_i\) 
such that \(\rho_L (n_i) \ge b_0\, n_i^l\).

\begin{Lem}\label{L1} If \(L\) is a regular language such that \(\rho_L (n)\) is 
 \(\Theta (n^l)\) for some integer \(l\) then there exist constants 
\(b_0\) and \(C\) and an  
infinite sequence \(n_0\), \(n_1\), \(\ldots\) \(,n_i,\ldots \) such 
that for all \(i\in \N\), \(\rho_L (n_i) \ge b_0\, n_i^l\) and 
\(n_{i+1}-n_{i}=C\). 
\end{Lem}

\begin{proof}It is obvious that there exists a word \(w \in L\) which 
is \((l+1)\)-tiered (see \cite[Lemmas 2-4]{Yu}), \(w=x\, y_1^{d_1}\, z_1 \ldots  y_{l+1}^{d_{l+1}}\, 
z_{l+1}\). Let \(C=|y_1|\ldots |y_{l+1}|\). As shown in \cite{Yu}, 
there exists a constant \(b_0\) such that the 
number of words of length \(n_t=|x z_1 \ldots z_{l+1}|+t\, C\) is 
greater than \(b_0\, n_t^l\) for any integer \(t\). 
\end{proof}

Recall (see \cite{Bo}) that the finite sum of integral powers is given by
\[\sum_{i=0}^n i^p=\frac{(n+B+1)^{p+1}-B^{p+1}}{p+1}\]
where all terms of the form \(B^m\) are replaced with the 
corresponding Bernoulli numbers \(B_m\). This formula will be useful 
in the next lemma.

\begin{Lem}\label{L2} If \(\rho_L (n)\) is \(\Theta (n^l)\) then 
\(v_n=\sum_{i=0}^n \rho_L (i)\) is \(\Theta (n^{l+1})\). Moreover, 
there exists a constant \(J\) such that \(v_{n_{i}}\ge J\, n_i^{l+1}\) 
for the sequence \(n_0,n_1,\ldots ,n_i,\ldots \) of Lemma \ref{L1}.
\end{Lem}

\begin{proof} i) There exist \(N_0\) and a constant \(b_1\) 
such that for all \(n\ge N_0\), 
\(\rho_L (n)\le b_1\, n^l\). If one replaces \(b_1\) by a bigger 
constant then the latter inequality holds for all \(n\). For 
\(n\) sufficiently large,  there exists a constant \(K\) such that
\[v_n=\sum_{i=0}^n \rho_L (i)\le b_1 \sum_{i=0}^n i^l \le K \, 
n^{l+1}.\]

ii) With the sequence \(n_i\) of Lemma \ref{L1}, one has
\[v_{n_{i}}=\sum_{j=0}^{n_i} \rho_L (j) \ge \sum_{j=0}^{i} \rho_L 
(n_j)\ge b_0 \, \sum_{j=0}^{i} (n_0+j\, C)^l \ge b_0 \,C^l 
\sum_{j=0}^{i} j^l.\]
Since \(n_i=n_0+i\, C\), then \(n_i\) is a linear function of \(i\) 
and for \(i\) large enough, there exists a constant \(J\) such that
\[v_{n_i}\ge J\, n_i^{l+1}.\]
\end{proof}

So, at this stage, we have a sequence \(n_i\) such 
that \(n_i=n_0+i\, C\) and constants \(b_0\), \(b_1\), \(K\) and 
\(J\) such that for \(n\) and \(i\) sufficiently large,
\[\left\{\begin{array}{ll}
\rho_L (n) &\le b_1\, n^l \cr
\rho_L (n_i) & \ge b_0\, n_i^l \cr \end{array}\right. \ {\rm and}\ 
\left\{\begin{array}{ll}
v_n &\le K\, n^{l+1} \cr
v_{n_i} & \ge J\, n_i^{l+1} \cr \end{array}\right. .
\]

Before going further in the proof of Theorem \ref{Grand}, we give an 
interesting result about the convergence of the sequence 
\((\frac{v_n}{n^{l+1}})_{n\in \N}\) when \(L\) is a polynomial language. 
A remarkable fact is that the limit always 
exists. Although this is generally not the case for the sequence 
\((\frac{\rho_L(n)}{n^l})_{n\in \N}\). Consider for instance the language 
\(W=a^*b^*\cap(\{a,b\}^2)^*\). It is obvious that 
\(\rho_W(2n+1)=2n+2\), \(\rho_W (2n)=0\) and \(v_{2n}=v_{2n+1}=(n+1)^2\).

\begin{Lem}\label{eit} Let \(\rho_1,\ldots 
,\rho_k,\theta_1,\ldots ,\theta_k,\Phi_1,\ldots ,\Phi_k\) be real 
numbers such that for all \(i\neq j\), \(\theta_i\neq \theta_j\) and 
for all \(j\), \(\rho_j\neq 0\). 
There exists \(\varepsilon >0\) such that
\[M_n=|\rho_1 \, e^{i(n\theta_1 +\Phi_1)}+\cdots +\rho_k \, e^{i(n\theta_k 
+\Phi_k)}|>\varepsilon\]
for an infinite sequence of integers \(n\).
\end{Lem}

\begin{proof} Assume that for all \(\varepsilon>0\), 
\(M_n\ge\varepsilon\) only for a finite number of integers \(n\). In 
other words, \(M_n\to 0\). By successive applications of 
Bolzano-Weierstrass'theorem, there 
exist complex numbers \(z_1,\ldots ,z_k\) and a subsequence \(k(n)\) 
such that
\[\rho_j \, e^{i\, (k(n)\, \theta_j +\Phi_j)}\to z_j \ {\rm and}\ |z_j|=\rho_j\neq 0.\]
Since \(M_n\to 0\), then \(\sum_{j=1}^k z_j=0\). For \(l=0,\ldots ,k-1\), 
one gets in the same manner 
\[\sum_{j=1}^k \rho_j \, e^{i\, [(k(n)+l)\, \theta_j +\Phi_j]} \to \sum_{j=1}^k 
z_j \, e^{i\, l\, \theta_j}=0.\]
Therefore one has
\[\left(\begin{array}{cccc}
1 & 1 & \ldots & 1 \\ 
e^{i\theta_1} & e^{i\theta_2} & \ldots & e^{i\theta_k} \\ 
\vdots & \vdots & & \vdots \\
e^{i \, (k-1)\theta_1} & e^{i\, (k-1)\theta_2} & \ldots & e^{i\, (k-1)\theta_k} 
\end{array}\right)
\left(\begin{array}{c}
z_1 \\ z_2\\ \vdots \\ z_k \\ \end{array}\right) = 
\left(\begin{array}{c}
0 \\ 0\\ \vdots \\ 0 \\ \end{array}\right) .\]
This equality leads to a contradiction since the Vandermonde 
determinant does not vanish. 
\end{proof}

We are now able to prove the convergence of 
\(({v_n}/{n^{l+1}})_{n\in\N}\). This result and its proof were suggested by P.~Lecomte.

\begin{The}\label{struct} If \(L\) is a regular language such that 
\(\rho_L (n)\) is \(\Theta (n^l)\) then the sequence 
\((\frac{v_n}{n^{l+1}})_{n\in\N}\) converges to a strictly positive limit. 
Moreover, \(1\) is a root of the characteristic polynomial of the 
sequence \((\rho_{L}(n))_{n\in\N}\) with a multiplicity equal to \(l+1\).
\end{The}

\begin{proof} The sequence \((\rho_L (n))_{n\in \N}\) satisfies a 
recurrence relation. Therefore, if \(z_i\) is a root of multiplicity 
\(\alpha_i\) of the characteristic polynomial of \((\rho_L(n))_{n\in 
\N}\) then one can write
\begin{equation}\label{pzz}
\rho_L (n)=\sum_{i} P_i (n) \, z_i^n
\end{equation}
where \(P_i (n)\) is a polynomial of degree less than 
\(\alpha_i\). Moreover \(\rho_L (n)\) is \(\Theta (n^l)\); in other 
words, we have a 
constant \(K\) such that
\begin{equation}\label{inegK}
\frac{\rho_L(n)}{n^l}\le K.
\end{equation}
This latter inequality has important consequences.

i) We first show that \(|z_i|>1\) implies \(P_i =0\). Otherwise, let 
\(\tau=\sup_{i} |z_i|\) and \(d\) the maximal degree of  polynomials 
\(P_i\) corresponding to the different roots of modulus \(\tau\). So we can write
\[\left\vert \frac{\rho_L(n)}{n^l} \right\vert = \frac{\tau^n \, n^d}{n^l} 
\, \vert c_1\, 
e^{in\theta_1} + \cdots + c_t\, e^{in\theta_t} + R_n \vert .\]
In the last expression, \( R_n \) is made up of two sorts  of terms, 
namely
\[R_n=\frac{1}{\tau^n \, n^d}\left( \sum_{j:|z_j|<\tau} P_j(n) z_j^n + 
\sum_{j:|z_j|=\tau} 
h_d(P_j(n)) z_j^n \right) \]
where \[ h_d:\C [z]\to \C [z] : P(z) \mapsto P(z)-\frac{D_z^d P(z)}{d!} \, z^d.\]
So \(R_n\to 0\) if \(n\to +\infty\). Therefore, by Lemma \ref{eit}, 
there exists an infinite sequence of integers such that
\[\left\vert \frac{\rho_L(n)}{n^l} \right\vert \ge \frac{\tau^n \, n^d}{n^l} (\varepsilon - 
\vert R_n \vert ).\]
For \(n\) large enough, \(\vert R_n \vert \le \varepsilon/2\) and 
\(\vert \frac{\rho_L(n)}{n^l} \vert \ge 
\tau^n \, n^{d-l} \, \frac{\varepsilon}{2}\) occurs infinitely often which contradicts 
(\ref{inegK}).

\medskip
ii) In the same way, one can verify that if \(|z_i|=1\) then the 
degree of the corresponding polynomial \(P_i\) cannot exceed \(l\).

\medskip
iii) If we are interested in the behaviour of \(v_n/n^{l+1}\) when \(n\to 
+\infty\), then in the expression (\ref{pzz}), we simply focus 
on the terms of the form \(\frac{D_n^l P_i(n)}{l!} n^l\, z_i^n\) for \(i\) 
such that \(|z_i|=1\). 
Indeed, any other term in \(\rho_L (n)\) provides \(v_n/n^{l+1}\) with 
a term which converges to zero (all these terms are included in 
\(R_n'\)).
So, if 
we assume that \(1=z_0\) has a multiplicty \(l+1\) and if 
\(z_1=e^{i\theta_1}\),\(\ldots\), \(z_t=e^{i\theta_t}\) are the other roots 
of modulus one with \(q_j=\frac{D_n^l P_j(n)}{l!}\), \(j=0,\ldots,t\) 
and \(\theta_0=0\); 
then one can write
\[\rho_L (n)  = \sum_{j=1}^t q_j\, n^l \, e^{in\theta_j} + q_0 \, 
n^l + R_n'\]
with \(\theta_j\neq 0\) and
\[R_n'=\sum_{j=0}^t h_l(P_j(n)) e^{in\theta_j} + \sum_{j:\vert 
z_j\vert<1} P_j(n) z_j^n.\]
Therefore, 
it is easy to see that 
\[\frac{v_n}{n^{l+1}}=  
\sum_{j=1}^t q_j\, \underbrace{\frac{1}{n^{l+1}}\, \sum_{k=0}^n k^l \, 
e^{ik\theta_j}}_{\to 0} + q_0 \, \underbrace{\frac{1}{n^{l+1}}\, 
\sum_{k=0}^n k^l}_{\to 1/l+1} + \underbrace{\frac{1}{n^{l+1}}\, 
\sum_{k=0}^n R_k'}_{\to 0}.\]
Moreover, we see that \(1\) has, necessary, a multiplicity \(l+1\); 
otherwise, \(\frac{v_n}{n^{l+1}} \to 0\), which is a contradiction 
with Lemma \ref{L2}.
\end{proof}


\medskip
\noindent
{\it Proof of Theorem \ref{Grand}.} By definition of a numeration 
system, it is clear that for \(n\) sufficiently large,
\( n+1 \le |\r{S}(v_n)| \le n+C+1\) since for 
\(C\) consecutive values of \(\rho_L (n)\) at least one of them does not 
vanish. (Notice that if \(\rho_L (n)>0\) for all \(n\), then 
\(|\r{S}(v_n)|=n+1\).) Recall also 
that \(|\r{S}(x)|=n\) iff \(v_{n-1}\le x< v_n\).
In this proof, we use the sequence \(n_i\) and the constants \(J\), 
\(K\), \(b_0\) and \(b_1\) introduced in the previous propositions.

\medskip
i) Assume that the integer constant 
\(\lambda\) is strictly greater than \(\left(\frac{K}{J}\right)^l\). We 
show that for \(n\) large enough,
\begin{equation}\label{inegceil}
n+1\le |\r{S}(\lambda \, v_n)| \le \lceil \lambda^{1/l}\, n \rceil + C 
-1 < \lambda^{1/l}\, n +C. 
\end{equation}
It is sufficient to show that \(\lambda \, v_n < v_{\lceil \lambda^{1/l}\, n 
\rceil + C -1 }\). By Lemma \ref{L2}, there exists \(k\in \{ \lceil \lambda^{1/l}\, n \rceil , 
\ldots , \lceil \lambda^{1/l}\, n \rceil +C-1\}\) such that \(v_k\ge 
J\, k^{l+1}\). Moreover the function \(n\mapsto v_n\) is increasing. 
So, 
\[ v_{\lceil \lambda^{1/l}\, n \rceil + C -1 } \ge J\, \lceil 
\lambda^{1/l}\, n \rceil^{l+1} \ge J\, \lambda^{\frac{l+1}{l}} \, n^{l+1}.\]
Moreover, by Lemma \ref{L2}, \(\lambda\, v_n\le \lambda\, K\, 
n^{l+1}\). By the choice of \(\lambda\), it is clear that \(\lambda\, K\, 
n^{l+1} <  J\, \lambda^{\frac{l+1}{l}} \, n^{l+1}\).

\medskip
ii) In Lemma \ref{L1} and Lemma \ref{L2}, we have introduced two 
constants \(b_0\) and \(b_1\) such that \(b_0\le b_1\). 

Let \(s\in \N\backslash \{0\}\) such that 
\(s\, b_0>b_1\). Here, we show that the function
\[i \mapsto |\r{S}(\lambda\, v_{n_{si}-1})|\]
is strictly increasing for \(i\) sufficiently large. So, we have to show that
\[|\r{S}(\lambda\, v_{n_{s(i+1)}-1})|=|\r{S}(\lambda\, v_{n_{si}+sC-1})|
>|\r{S}(\lambda\, v_{n_{si}-1})|. \]
Let \(k=|\r{S}(\lambda\, v_{n_{si}-1})|\) then \(v_{k-1}\le \lambda\, 
v_{n_{si}-1} < v_k\) and we must show that 
\[\lambda\, v_{n_{si}+sC-1} = \lambda\, v_{n_{si}-1} + \lambda\, 
\sum_{j=0}^{sC-1} \rho_L (n_{si}+j) \ge v_k=v_{k-1}+\rho_{L}(k).\]
So, it is sufficient to show that \(\lambda\, 
\sum_{j=0}^{sC-1} \rho_L (n_{si}+j) \ge \rho_{L}(k)\). 
In view of (\ref{inegceil}), \(k<\lambda^{1/l} (n_{si}-1) + C\). 
Therefore \(\rho_{L}(k)< b_1\, [\lambda^{1/l} (n_{si}-1) + C]^l\). On the 
other hand, 
\[ \lambda\, \sum_{j=0}^{sC-1} \rho_L (n_{si}+j) \ge 
\lambda\, \sum_{j=0}^{s-1} \underbrace{\rho_L (n_{si}+j\, C)}_{\ge b_0\, 
(n_{si}+j\, C)^l} \ge \lambda\, b_0 \, s\, n_{si}^l.\]
To conclude this part, notice that the coefficient of \(n_{si}^l\) in 
\(b_1\, [\lambda^{1/l} (n_{si}-1) + C]^l\) is \(b_1\, \lambda\) and by 
choice of \(s\), we have 
\( b_1\, \lambda < \lambda\, b_0 \, s\). So the inequality holds for 
\(i\) sufficiently large.

\medskip
iii) Consider the subset
\[X=\{ v_{n_{si}-1} : i \in \N\}= \{ v_{n_0+siC-1} : i \in \N\}.\]
Since \(\rho_{L} (n_0+siC)>0\), then \(\r{S}(v_{n_0+siC-1})\) is the first 
word of length \(n_0+siC\) and
\[\r{S} (X)=\r{S}\left(\{ v_n : n\in \N\} \right) \cap \Sigma^{n_0} 
\left(\Sigma^{sC} \right)^*.\] 
So \(X\) is a \(S\)-recognizable subset of \(\N\) \cite{Sh}.

Assume that \(\lambda\, X\) is recognizable. Therefore, \(|\r{S}(\lambda\, 
X)|\) is a finite union of arithmetic progressions. In view of ii), we 
can apply Lemma \ref{fctcr} and obtain two integral constants 
\(\Gamma\) and \(k\) such that for all \(\alpha \in \N\),
\[|\r{S}(\lambda\, v_{n_0+sC(i+\alpha\, k)-1})|=|\r{S}(\lambda\, 
v_{n_0+sCi-1})|+\alpha\, \Gamma.\]
Or equivalently, if we set \(z=|\r{S}(\lambda\, v_{n_0+sCi-1})|\) then
\begin{equation}\label{llg}
v_{z+\alpha\, \Gamma-1}\le \lambda\, v_{n_0+sC(i+\alpha\, k)-1} < 
v_{z+\alpha\, \Gamma}.
\end{equation}
First consider the left inequality in (\ref{llg}), 
with the same argument as in i), we obtain 
\[v_{z+\alpha\, \Gamma-1}\ge J\, (z+\alpha\, \Gamma-C)^{l+1}.\]
On the other hand, 
\[\lambda\, v_{n_0+sC(i+\alpha\, k)-1} \le \lambda\, K\, (n_0+sCi+sCk\, \alpha 
-1)^{l+1}.\]
Since \(\alpha\) can be arbitrary large, we focus on the terms of the form \(\alpha^{l+1}\).
Then we obtain the following condition, 
\begin{equation}\label{cond1}
J\, \Gamma^{l+1}\le \lambda\, K \, (sCk)^{l+1} \ {\rm or}\ 
\lambda\ge \frac{J}{K}\, \left( \frac{\Gamma}{sCk} \right)^{l+1}.
\end{equation}
If we consider the right inequality in (\ref{llg}), we have 
\(v_{z+\alpha\, \Gamma}\le K\, (z+\alpha\, \Gamma)^{l+1}\) and also
\[\lambda\, v_{n_0+sC(i+\alpha\, k)-1}\ge \lambda\, J\, (n_0+sCi+sCk\, \alpha 
-C)^{l+1}.\]
If we focus on terms in \(\alpha^{l+1}\), we obtain
\begin{equation}\label{cond2}
\lambda\le \frac{K}{J}\, \left( \frac{\Gamma}{sCk} \right)^{l+1}.
\end{equation}

\medskip
iv) By Theorem \ref{struct}, \((\frac{v_n}{n^{l+1}})_{n\in\N}\) converges to a 
limit \(a>0\). Consider the sequences
\[K_m=a+\frac{1}{m} \ {\rm and}\ J_m=a-\frac{1}{m}.\]
For a given \(m\) there exist \(i_m\) and \(n_m\)
such that for \(i\ge i_m\), \(v_{n_i}\ge J_m\, n_i^{l+1}\) and 
for \(n\ge n_m\), \(v_n\le K_m\, n^{l+1}\). So, if we replace \(K\) 
by \(K_m\) and \(J\) by \(J_m\), the previous points i), ii) and iii) 
remain true for \(n\) sufficiently large. 

For \(m\) large enough, the condition \(\lambda > 
\left(\frac{K_m}{J_m}\right)^l\) given in i) is equivalent to \(\lambda\ge 2\) and the 
conditions (\ref{cond1}) and (\ref{cond2}) may be replaced by a unique 
condition
\[\lambda=\left( \frac{\Gamma}{sCk} \right)^{l+1}\]
which contradicts the hypothesis (remember that \(\Gamma,s,C\) and 
\(k\) are integers). \(\Box\)

\section{Multiplication and complement of polynomial languages}\label{NS5}    

In the previous sections, we have considered multiplication for  
numeration systems based on a polynomial language. If the complexity 
function of a regular language is not bounded by a polynomial then it is of order 
\(2^{\Theta(n)}\) and the language is said to be {\it exponential}. 
The class of exponential languages 
splits into two subclasses according whether the complement of a 
language is polynomial or not.

In this section, we have a closer look at 
numeration systems constructed on an exponential regular language 
such that its complement has a complexity function bounded by a 
polynomial. We show that for such systems, multiplication 
by a constant generally does not preserve recognizability.

We begin with the example of \(\Sigma^*\setminus L\) 
where \(L\) is the polynomial language \(a^* b^*\) and 
\(\Sigma=\{a,b\}\). Thus, with \(S=(\Sigma^*\setminus 
L,\{a,b\},a<b)\), we compute the representations of \(2\, v_n\) and 
obtain Table \ref{tab} (for an algorithm of representation, see 
\cite{LR}).

\begin{table}[htbp]\label{tab}    
\begin{center}
\begin{tabular}{| c | c || r l | c | c |}
\hline
\(n\) & \(2\, v_n\) & \(\r{S}(2\, v_n)=b^k\) & \(\!\!\!\!\!\! aw\) & \(k\) & \(|w|\) \\
\hline
1 & 0 & \(b\) &  & 1& 0\\
2 & 2 & \(b\) & \(\!\!\!\!\!\! aa\) & 1& 1\\
3 &10 & \(b\) & \(\!\!\!\!\!\! aab\) & 1& 2\\
4 & 32 & \(b\) & \(\!\!\!\!\!\! abab\) & 1& 3\\
5 & 84 & \(bb\) & \(\!\!\!\!\!\! aaaa\) & 2& 3\\
6 & 198 & \(bb\) & \(\!\!\!\!\!\! ababa\) & 2& 4\\
7 & 438 & \(bbb\) & \(\!\!\!\!\!\! aaabb\) & 3& 4\\
8 & 932 & \(bbb\) & \(\!\!\!\!\!\! abbabb\) & 3& 5\\
9 & 1936 & \(bbbb\) & \(\!\!\!\!\!\! abaaba\) & 4& 5\\
10 & 3962 & \(bbbbb\) & \(\!\!\!\!\!\! aabaaa\) & 5& 5\\
11 & 8034 & \(bbbbb\) & \(\!\!\!\!\!\! abbbbab\) & 5& 6\\
12 & 16200 & \(bbbbbb\) & \(\!\!\!\!\!\! abbaaab\) & 6& 6\\
13 & 32556 & \(bbbbbbb\) & \(\!\!\!\!\!\! abaabaa\) & 7& 6\\
14 & 65294 & \(bbbbbbbb\) & \(\!\!\!\!\!\! aababba\) & 8& 6\\
15 & 130798 & \(bbbbbbbbb\) & \(\!\!\!\!\!\! aaaabbb\) & 9& 6\\
16 & 261836 & \(bbbbbbbbb\) & \(\!\!\!\!\!\! abbbabbb\) & 9& 7\\
17 & 523944 & \(bbbbbbbbbb\) & \(\!\!\!\!\!\! abbaabba\) & 10& 7\\
18 & 1048194 & \(bbbbbbbbbbb\) & \(\!\!\!\!\!\! abababaa\) & 11& 7\\
19 & 2096730 & \(bbbbbbbbbbbb\) & \(\!\!\!\!\!\! abaaaaab\) & 12& 7\\
20 & 4193840 & \(bbbbbbbbbbbbb\) & \(\!\!\!\!\!\! aababbab\) & 13& 7\\
21 & 8388100 & \(bbbbbbbbbbbbbb\) & \(\!\!\!\!\!\! aaabbaaa\) & 14& 7\\
 \hline
\end{tabular}
\caption{first terms of \(2\, v_n\) for \(S=(\{a,b\}^*\setminus 
a^*b^*,\{a,b\},a<b)\).}
\end{center}
\end{table}
In view of this table, it appears that the number of leading 
\(b\)'s in the representation is increasing. Furthermore, it seems that the 
length of the tail also increases. Let us show that this observation 
is true and can be generalized.

\begin{Def}{\rm Let \(L\subset \Sigma^*\) and \(x\in 
\Sigma^*\), we set \(L_x = \{w \in L : w=xy\}\). 
It is clear that \(L_x \subseteq L\). So \(\rho_{L_x} (n)\le \rho_L 
(n)\) and  \(\rho_{L_x}\) is  
\(O(n^l)\) whenever \(\rho_L\) is \(O(n^l)\). }\end{Def}

In our example, for \(0\le k< n\), we have
\[\rho_{(\Sigma^*\setminus L)_{b^{n-k}}} (n)=\rho_{\Sigma^*_{b^{n-k}}} 
(n) -\rho_{L_{b^{n-k}}} (n) = 2^k -1.\]
The complexity function \(\rho_{(\Sigma^*\setminus L)} (n)\) 
of the language associated to the system \(S\) is \(2^n-n-1\). 
So the sequence \(v_n\) associated to \(\Sigma^*\setminus L\) is 
\[ v_n=\sum_{i=0}^n  
\rho_{(\Sigma^*\setminus L)} (i)=2^{n+1}-\frac{n(n+3)}{2}-2.\]
The words of \(\r{S}(\{v_n:n\in \N\})\) are the first words of each 
length in \(\Sigma^*\setminus L\). So \(\{v_n:n\in \N\}\) is 
\(S\)-recognizable. Recall that 
\(\vert \r{S}(x)\vert = n \Leftrightarrow v_{n-1} \le x < v_n\).
For \(n\) large enough, it is obvious that 
\(v_n \le 2\, v_n < v_{n+1}.\) Then \(\vert \r{S}(2\, v_n)\vert=n+1\).

Let us show that \(\{2\, v_n:n\in \N\}\) is not \(S\)-recognizable. 
For each \(n\) there exists a unique \(i\) such that
\[\rho_{(\Sigma^*\setminus L)_{b^{n-i+1}}} (n)=2^{i-1}-1
<\underbrace{v_{n+1}-2\, v_n}_{=n(n+1)/2}\le2^i-1=
\rho_{(\Sigma^*\setminus L)_{b^{n-i}}} (n).\]
Then \(\r{S}(2\, v_n)=b^{n-i}az\) with \(\vert z\vert=i\). 
Notice that, as a function of \(n\), \(i\) is increasing but grows more 
slowly than \(n\) (in fact, it has a logarithmic growth). So \(n-i\to +\infty\).

Assume that \({\cal L}=\r{S}(\{2\, v_n:n\in \N\})\) is accepted by an automaton 
with \(q\) states. There exist \(n_0\), \(i_0\) and \(t\ge 0\) such that 
\(\r{S}(2\, v_{n_0}) = 
b^{q+t} a z_0\) with \(\vert z_0\vert=i_0\). By the pumping lemma, there 
exists \(\alpha>0\) such that \[\forall m\in \N,\  
b^{q+t+m\alpha} a z_0 \in {\cal L}.\] In this last 
expression, \(z_0\) has a constant length \(i_0\) independent of \(m\). 
A contradiction.

\medskip
In view of this example, we state the following theorem. Recall that 
the complexity of any polynomial language is \(\Theta(n^l)\) for some 
\(l\). 
\begin{The}\label{thepol} Let \(\Sigma=\{\sigma_1< \cdots <\sigma_{s-1}<\beta\}\), 
\(s\ge 2\) and 
\(L\subset \Sigma^*\) be a regular language such that \(\rho_L (n)\) 
is \(\Theta (n^l)\). If \(S=(\Sigma^*\setminus L,\Sigma,<)\) then
there exists an \(S\)-recognizable set \(X\subset \N\) such that for 
all \(j\ge 1\), \(s^j X\) is not \(S\)-recognizable.
\end{The}
\begin{proof} For \(0\le k<n\), we have 
 \[\rho_{(\Sigma^*\setminus L)_{\beta^{n-k}}} 
 (n)=\rho_{\Sigma^*_{\beta^{n-k}}} 
(n) -\rho_{L_{\beta^{n-k}}} (n) = s^k -\underbrace{\rho_{L_{\beta^{n-k}}} 
(n)}_{\in O(n^l)}.\]
To avoid any misunderstanding, \(v_n\) is the sequence associated 
to   
the language \(\Sigma^*\setminus L\) of the numeration \(S\) and \(v_n(L)\) is 
related to \(L\). So, 
\(v_L(n)=\sum_{i=0}^n \rho_{L}(i)\) and
\[ v_n=\sum_{i=0}^n  
\rho_{(\Sigma^*\setminus L)} (i) = \frac{s^{n+1}-1}{s-1}-v_L(n).\]
We take \(X=\r{S}(\{v_n:n\in \N\})\), an \(S\)-recognizable set. We 
have, for \(n\) sufficiently large,  
\[v_{n+j-1}\le s^j v_n<v_{n+j}.\]
Indeed, \(v_{n+j}-s^j\, v_n=s^j \, v_L (n)-v_L 
(n+j)+\frac{s^j-1}{s-1}\). 
By Theorem \ref{struct}, there exists \(a>0\) such that \(v_L(n) \sim a\, 
n^{l+1}\). So \(v_{n+j}-s^j\, v_n\sim (s^j-1)a\, n^{l+1}\). On the 
other hand,  
\(s^j\, v_n-v_{n+j-1}=s^{n+j}+v_L (n+j-1)-s^j \, v_L 
(n)-\frac{s^j-1}{s-1}\) has an exponential dominant term. Then \(\vert 
\r{S}(s^j\, v_n)\vert=n+j\). 

For all \(n\) sufficiently large, there 
exists a unique \(i\) such that
\begin{equation}\label{coincage}
\underbrace{\rho_{(\Sigma^*\setminus L)_{\beta^{n-i+1}}} (n)}_{= s^{i-1} 
-\rho_{L_{\beta^{n-i+1}}} (n)}
<v_{n+j}-s^j\, v_n\le \underbrace{\rho_{(\Sigma^*\setminus L)_{\beta^{n-i}}} (n)}_{= s^{i} 
-\rho_{L_{\beta^{n-i}}} (n)}
\end{equation}
Then \(\r{S}(s^j\, v_n)=\beta^{n-i} \sigma z\) with \(\vert 
z\vert=i+j-1\) 
and \(\sigma\neq\beta\). 
Notice that as a function of \(n\), \(i\) is increasing and not 
bounded. To show that \(n-i\to 
+\infty\) if \(n\to +\infty\). Assume that \(n-i\) is bounded,  
divide all members of (\ref{coincage}) by \(s^n\). 
Let \(n\to +\infty\) and obtain a contradiction.

Suppose that \(\r{S}(\{s^j X\})\) is accepted by an automaton 
with \(q\) states. There exist \(n_0\), \(i_0\) and \(t\ge 0\) such that 
\(\r{S}(s^j v_{n_0}) = 
\beta^{q+t} \sigma z_0\) with \(\vert z_0\vert=i_0\) and 
\(\sigma\neq\beta\). Then using the pumping lemma, we obtain a contradiction.
\end{proof}

\section{Relation with positional numeration systems}\label{NS6}

In this section, we give sufficient conditions to achieve the 
computation of an \(U\)-representation of an integer from its  
\(S\)-representation, where \(U\) is some positional numeration 
system related to a sequence of integers. In particular, we obtain 
sufficient conditions to guarantee the stability of the 
\(S\)-recognizability after addition and multiplication by a constant.

Let us recall some definitions. A {\it \(2\)-tape automaton} over 
\(A^*\times B^*\) (also called {\it transducer}) 
is a directed graph with edges labelled by elements 
of \(A^*\times B^*\). The automaton is finite if the set of edges is 
finite.  A \(2\)-tape automaton is said {\it letter-to-letter} if  
the edges are labelled by elements of \(A\times B\). A relation 
\(R\subset A^*\times B^*\) is 
said to be {\it computable by a finite \(2\)-tape automaton} if there 
exists a finite \(2\)-tape automaton over 
\(A^*\times B^*\) such that the set of labels of paths starting in an 
initial state and ending in a final state is equal to \(R\). Finally, 
a function is computable by a finite \(2\)-tape automaton if its 
graph is computable by a finite \(2\)-tape automaton.

\begin{Def}{\rm
If \(U=(U_n)_{n\in \N}\) is a sequence of integers and \(x=x_n\ldots 
x_0\), a word over an alphabet \(B\subset \Z\). We define the 
{\it numerical value} of \(x\) as
\[\pi_U (x) = \sum_{i=0}^n x_i\, U_i.\]
Notice that different words can have the same numerical value.
}\end{Def}

\begin{Pro}\label{expexp} Let \(L\subset \Sigma^*\) be a regular 
language,  
\(M=(K,s,F,\Sigma,\delta)\) be a DFA accepting \(L\) and 
\(S=(L,\Sigma,<)\). Let 
\(U=(U_n)_{n\in \N}\) be a sequence of integers such that \(U_0=1\). 
If there exist \(k,\alpha \in \N\setminus\{0\}\), \(e_{p,i} \in \Z\) 
(\(p\in K\), \(i=0,\ldots ,k-1\)) such that for all state \(p 
\in K\) and all \(n\in \N\)
\begin{equation}\label{decomp}
\alpha\, u_{n+k-1} (p)= \sum_{i=0}^{k-1} e_{p,i}\, U_{n+i}.
\end{equation}
Then there exist a finite alphabet \(B\subset \Z\) and a finite 
letter-to-letter automaton which compute a function \(g:L\to B^*\) 
such that \(\vert w \vert = \vert g(w)\vert\) and \[\alpha\,  
\val{S}(w)=\pi_U (g(w)).\]
\end{Pro}

\begin{Rem}{\rm
The function \(g\) of the previous theorem is injective. If \(v\) and 
\(w\) are two words of \(L\) such that \(g(v)=g(w)\) then 
\(\val{S}(v)=\val{S}(w)\). So the conclusion, since \(\val{S}\) is 
a one-to-one correspondence.}
\end{Rem}

\begin{proof} We consider words of length at least \(k\). Indeed, there is 
only a finite number of words of length less than \(k\) and they can be 
treated separately. 
Let \(w=w_{k+l}\ldots w_{k-1} w_{k-2}\ldots w_0\) be a word of \(L\) of 
length \(k+l+1\) with \(l\ge-1\). We compute \(l+2\) applications of 
Lemma \ref{lemme} on \(\val{s} (w)\) and we obtain
\[\begin{array}{c}
{\displaystyle \sum_{\sigma<w_{k+l}}} u_{k+l}(s.\sigma) + 
{\displaystyle \sum_{i=-1}^{l}} u_{k+i} (s) + 
{\displaystyle \sum_{i=-1}^{l-1}}\   
{\displaystyle \sum_{\sigma<w_{k+i}}} u_{k+i} (s.w_{k+l}\ldots 
w_{k+i+1}\sigma) \cr 
+ \val{s.w_{k+l}\ldots w_{k-1}}(w_{k-2}\ldots w_0) + v_{k-2} (s) 
- v_{k-2} (s.w_{k+l}\ldots w_{k-1}).\cr \end{array}\]
Recall that the notation \(p.\sigma\) 
is written in place of \(\delta(p,\sigma)\). We will denote by \(C_w\) 
the sum of the last three terms. For all \(q \in K\), \(p 
\in K\setminus \{s\}\) and \(\sigma \in \Sigma\), let us define
\[\beta_{q,p,\sigma} = \# \{\sigma' < \sigma : q.\sigma' = p\}\]
and 
\[\beta_{q,s,\sigma} = 1+ \# \{\sigma' < \sigma : q.\sigma' = s\}.\]
With these notations, we can rewrite \(\val{s} (w)\) as 
\[C_w+{\displaystyle \sum_{p\in K}} \beta_{s,p,w_{k+l}}\, u_{k+l}(p) + 
{\displaystyle \sum_{i=-1}^{l-1}  \sum_{p\in K}} 
\beta_{s.w_{k+l}\ldots w_{k+i-1},p,w_{k+i}}\, u_{k+i}(p).\]
Therefore, using (\ref{decomp}), we have
\begin{eqnarray*}
\alpha\, \val{s}(w)=&\alpha \, C_w + {\displaystyle 
\sum_{j=0}^{k-1}} 
\underbrace{{\displaystyle \sum_{p\in K}} \beta_{s,p,w_{k+l}}\, 
e_{p,j}}_{=\lambda_{l,j}} \, U_{l+j+1}\\
 & + {\displaystyle \sum_{i=-1}^{l-1} \sum_{j=0}^{k-1}} 
\underbrace{{\displaystyle \sum_{p\in K}} \beta_{s.w_{k+l}\ldots w_{k+i-1},p,w_{k+i}}\, 
e_{p,j}}_{=\lambda_{i,j}} \, U_{i+j+1}.\end{eqnarray*}
It is obvious that the \(\lambda_{i,j}\)'s take their values in a 
finite set \(R\). Therefore sums of \(k-1\) elements of \(R\) also 
take their values in a finite set, say \(T\). 
Notice that the \(\lambda_{i,j}\)'s (resp. the \(\lambda_{l,j}\)'s) 
 are completely determined by the letter \(w_{k+i}\) (resp. \(w_{k+l}\)) and 
 the state \(s.w_{k+l}\ldots w_{k+i-1}\) reached after the lecture of the 
 first letters of \(w\) (resp. the state \(s\)). Therefore, we 
 extend the notation \(\lambda_{i,j}\) to a meaningful one:
\begin{equation}\label{deflambda}
\lambda_{q,\sigma,j} ={\displaystyle \sum_{p\in K}} \beta_{q,p,\sigma}\, 
e_{p,j} \end{equation}
with \(q\in K\), \(\sigma \in \Sigma\) and \(j=0,\ldots,k-1\).

We are now able to build a finite letter-to-letter \(2\)-tape 
automaton \({\cal M}\) over \(\Sigma^* \times B^*\) with \(B\subset \Z\) 
some  
finite alphabet. The formula expressing \(\alpha \val{s}(w)\) 
can be interpreted in the following way. 
The reading of \(w_{k+i}\), \(l\le i \le -1\), provides the 
decomposition of \(\alpha\, \val{s}(w)\) with \(\lambda_{i,k-1}\,  
U_{k+i}\); \(\lambda_{i,k-2}\, U_{k+i-1}\); \(\ldots \); \(\lambda_{i,0}\, U_{i+1}\). 
The reading of \(w_{k+i}\) gives a coefficient \(\lambda_{i,k-1}\) 
for \(U_{k+i}\). The other \(k-1\) coefficients can be viewed as ``remainders". 
Roughly speaking, if we have already read the word \(t=w_{k+l}\ldots 
w_{k+i+1}\) and if we are reading \(\sigma=w_{k+i}\),  
then we have to consider the state \(s.t\). (Therefore it seems 
natural to mimic \(M\) in \({\cal M}\).) The 
coefficients  \(\lambda_{i,k-1};\ldots ; \lambda_{i,0}\) are nothing 
else but \(\lambda_{s.t,\sigma,k-1};\ldots ; \lambda_{s.t,\sigma,0}\).

Thereby we can give a precise definition of \({\cal M}\). 
The set of states is 
\({\cal K}=K\cup\{f\} \times \underbrace{T \times \cdots \times T}_{k-1}\) 
where \(f\) does not belong to \(K\) and is the unique final state of 
\({\cal M}\). The copies 
of \(T\) will be used to store the ``remainders". The start state is 
\((s,0,\ldots ,0)\). The transition relation \(\Delta:{\cal K} \times 
(\Sigma \times B) \to {\cal K}\) is defined as 
follows. If \(p\in K\), \(\sigma \in \Sigma\),
\begin{eqnarray*}
 &\Delta((p,\gamma_{k-2},\ldots ,\gamma_0 
),(\sigma,\lambda_{p,\sigma,k-1}+\gamma_{k-2}))\\
=& (p.\sigma;
\lambda_{p,\sigma,k-2}+\gamma_{k-3};\ldots ; 
\lambda_{p,\sigma,1}+\gamma_0; 
\lambda_{p,\sigma,0})\end{eqnarray*}
These transitions compute an output \(x_{k+l}\ldots x_{k-1}\) from 
\(w_{k+l}\ldots w_{k-1}\). 
The alphabet \(B\) is finite since \(T\) is finite.

But we have still to read the last \(k-1\) 
letters of \(w\). For each state \(p\in K\), \(D_p=L_p \cap 
\Sigma^{k-1}\) is finite (recall that \(L_p\) are the words accepted 
from \(p\)). So, for each state \(p\in K\) and 
each word \(w_{k-2}\ldots w_0 \in D_p\), we construct an edge from  
\((p,\gamma_{k-2},\ldots ,\gamma_0)\) to \(f\) labelled by
\((w_{k-2}\ldots w_0, \gamma_{k-2}\cdots \gamma_1 (\gamma_0+C_w))\). 
(This kind of edge can naturally be split in \(k-1\) elementary 
edges using \(k-2\) new states.) 
Indeed, notice that \(C_w\) is a constant which only depends on the 
state \(s.w_{k+l}\ldots w_{k-1}\) reached (the first component in 
\({\cal K}\)) and the remainding word 
\(w_{k-2}\ldots w_0\).
\end{proof}

\begin{Rem}{\rm The states of \(M\) satisfy the same recurrence 
relation of degree \(l\). A practical way to check (\ref{decomp}) is to 
seek a final state \(f\in F\) such that
\[det\left(\begin{array}{ccc}
u_0(f) & \cdots & u_{l-1}(f) \cr
\vdots & & \vdots \cr
u_{l-1}(f) & \cdots & u_{2l-2}(f) \cr \end{array}\right)\neq 0.\]
If such an \(f\) exists then for all \(p \in K\), there exist 
\(c_{p,i} \in \Q\) such that
\[u_{n+l-1}(p)=\sum_{i=0}^{l-1} c_{p,i}\, u_{n+i}(f)\]
and (\ref{decomp}) can be easily obtained.
}\end{Rem}

Recall that a strictly increasing sequence \(U=(U_n)_{n\in \N}\) of 
integers such that \(U_0=1\) and \(\frac{U_{n+1}}{U_n}\) is bounded, 
defines a {\it positional numeration system}. If \(x\) is an integer, the 
{\it \(U\)-representation} of \(x\) obtained by the greedy algorithm is denoted 
by 
\(\rho_U (x)\) and belongs to \(A_U^*\) where \(A_U=\{0,\ldots ,Q\}\) is the {\it canonical 
alphabet} of the system \(U\), \(Q<\max \frac{U_{n+1}}{U_n}\). A set 
\(X\subset \N\) is said {\it\(U\)-recognizable} if \(\rho_U (X)\) is 
regular. For any alphabet \(C\) of integers, one 
can define a partial function called {\it normalization}
\[\nu_{U,C} : C^* \to A_U^* : z 
\mapsto \rho_U(\pi_U(z)).\] 

\begin{Cor}\label{premcor} Let \(S=(L,\Sigma,<)\). With the hypothesis and notations 
of  Proposition \ref{expexp}, if the sequence \(U\) defines a 
positional numeration system such that the normalization 
function \(\nu_{U,B}\) is computable by finite 
letter-to-letter \(2\)-tape automaton then \(X\subset \N\) is 
\(S\)-recognizable if and only if \(\alpha X\) is \(U\)-recognizable.
\end{Cor}

\begin{proof} Let the regular language \({\cal G}\subset (\Sigma \times B)^*\) 
be the graph of the function \(g\) defined in Proposition \ref{expexp}. 
We denote by \(p_1:\Sigma \times B \to \Sigma\) and 
\(p_2:\Sigma \times B \to B\) the canonical 
homomorphisms of projection. Let
\[Y=p_2[p_1^{-1} (\r{S}(X))\cap {\cal G}].\]
If \(X\) is \(S\)-recognizable then \(Y\subset B^*\) is regular and 
\(\pi_U (Y)=\alpha X\). So \(\alpha X\) is \(U\)-recognizable since 
\(\nu_{U,B}(Y)\) is regular.

Conversely, if \(\rho_U(\alpha X)\) is 
regular then \(\nu_{U,B}^{-1}\circ \rho_U(\alpha X)\) is also regular. For 
each \(y \in \alpha X\), \(\nu_{U,B}^{-1}\circ \rho_U(y)\) can take 
more than one value but only one is in \(p_2 ({\cal G})\). So the set
\[p_1\left(p_2^{-1}[\nu_{U,B}^{-1}\circ \rho_U(\alpha X)] \cap {\cal 
G}\right)\]
is regular and equal to \(\r{S}(X)\).
\end{proof}

\begin{Cor} Let \(S=(L,\Sigma,<)\). With the hypothesis and notations 
of  Proposition \ref{expexp}, if the sequence \(U\) satisfies a linear 
recurrence relation
\[U_n=d_1 U_{n-1} + \cdots + d_m U_{n-m}, d_i \in \Z, d_m \neq 0, n\ge 
m\]
such that its characteristic polynomial is the minimal polynomial of a 
Pisot number then  \(X\subset \N\) is 
\(S\)-recognizable if and only if \(X\) is \(U\)-recognizable.
\end{Cor}

\begin{proof} It is well known that for such a system \(U\) the 
normalization \(\nu_{U,C}\) is computable by finite 
letter-to-letter \(2\)-tape automaton for any alphabet \(C\) (see 
\cite{FS}). So by the previous corollary, \(X\) is \(S\)-recognizable if and only if 
\(\alpha X\) is \(U\)-recognizable. Another well-known fact related to 
Pisot numeration systems is that a subset \(X\) is \(U\)-recognizable 
if and only if it is definable in the structure \(\langle \N, +,V_U 
\rangle\) (see \cite{BH}). In particular, multiplication by a constant \(\alpha\) is 
definable in \(\langle \N , +\rangle \). So \(\alpha X\) is definable 
in the structure if and only if \(X\) is definable.
\end{proof}

\begin{Rem}{\rm Let \(S=(L,\Sigma,<)\) and \(S'=(L,\Sigma,\prec)\) be two 
systems which only differ by the ordering of the alphabet. If the 
hypothesis of Proposition \ref{expexp} and Corollary \ref{premcor} 
are satisfied then a set \(X\) is \(S\)-recognizable if and only if 
it is \(S'\)-recognizable. In other words, recognizable sets are 
independent of the ordering of the alphabet.
}\end{Rem}

\begin{Exa}{\rm Consider the language \(L\subset \{a,b,c\}^*\) of the 
words that do not contain \(aa\). Its minimal automaton \(M_L\) is given on 
Figure \ref{fig1}. As usual, the start state is indicated by an 
unlabeled arrow and the final states by double circles.
\begin{figure}[htbp]
 \begin{center}
\includegraphics{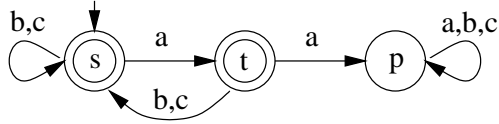}
 \end{center}
\caption{The minimal automaton of \(L\).}\label{fig1}
\end{figure}
The sequences associated to the different states satisfy the relation 
\[u_{n+2}=2u_{n+1}+2u_n, \forall n\in \N\]
with the initial conditions \(u_0 (s)=1\), \(u_1(s)=3\), \(u_0(t)=1\), 
\(u_1(t)=2\), \(u_0(p)=u_1(p)=0\). The sequence \(U\) of Proposition 
\ref{expexp} can be played by \((u_n(s))_{n\in \N}\). For all \(n\in \N\), 
we have the relations
\[\left\{\begin{array}{llll}
u_{n+1}(s)=1\, u_{n+1}(s) + 0\, u_n(s) &\Rightarrow &e_{s,0}=0, & e_{s,1}=1 \cr
u_{n+1}(t)=0\, u_{n+1}(s) + 2\, u_n(s) &\Rightarrow &e_{t,0}=2, & e_{t,1}=0 \cr
u_{n+1}(p)=0\, u_{n+1}(s) + 0\, u_n(s) &\Rightarrow &e_{p,0}=0, & e_{p,1}=0 \cr
\end{array}\right.\]
Notice that the characteristic polynomial of the recurrence satified 
by \(u_n(s)\) is \(x^2-2x-2=(x-1+\sqrt{3})(x-1-\sqrt{3})\). So 
\(U=(u_n(s))_{n\in \N}\) is a positional numeration system associated 
to the Pisot number \(1+\sqrt{3}\).  
From \(M_L\), we compute the \(3\times 3\) matrices 
\(B_{\sigma}=(\beta_{q,r,\sigma})_{q,r=s,t,p}\), \(\sigma \in \Sigma\) :
\[
B_a=\left(\begin{array}{ccc}1&0&0\cr 1&0&0\cr 1&0&0\cr 
\end{array}\right) , 
B_b=\left(\begin{array}{ccc}1&1&0\cr 1&0&1\cr 1&0&1\cr 
\end{array}\right) , 
B_c=\left(\begin{array}{ccc}2&1&0\cr 2&0&0\cr 1&1&2\cr 
\end{array}\right)
\]
If \(E=(e_{q,i})_{q=s,t,p;i=0,1}\) then it follows from 
(\ref{deflambda}) 
that \((B_\sigma E)_{q,i}=\lambda_{q,\sigma,i}\). We have
\[
B_aE=\left(\begin{array}{ccc}0&1\cr 0&1\cr 0&1\cr 
\end{array}\right) , 
B_bE=\left(\begin{array}{ccc}2&1\cr 0&1\cr 0&1\cr 
\end{array}\right) , 
B_cE=\left(\begin{array}{ccc}2&2\cr 0&2\cr 2&1\cr 
\end{array}\right)
\]
To obtain the complete transducer, with the notations of the proof of 
Proposition \ref{expexp}, we have to compute the \(C_w\) namely 
\[C_{q,\sigma}=\val{q}(\sigma)+v_0(s)-v_0(q)\]
for \(q\) and \(\sigma\) such that \(q.\sigma \in F\). Finally we 
have on Figure \ref{fig2} the finite letter-to-letter automaton build from 
\(M_L\) and the \(\lambda_{q,\sigma,i}\)'s. 
\begin{figure}[htbp]
 \begin{center}
\includegraphics{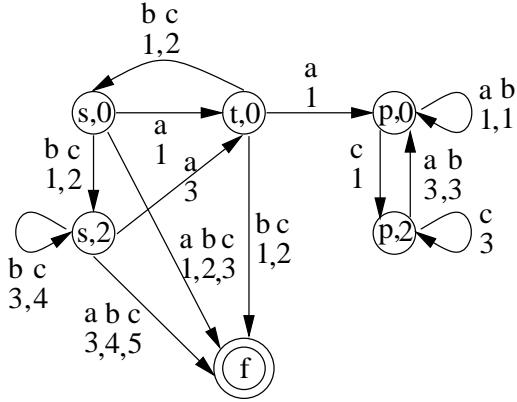}
 \end{center}
\caption{The transducer computing \(g\).}\label{fig2}
\end{figure}

We can do the same construction for the language \(L'=a^+\{a,b\}^*\). 
Its minimal automaton \(M_{L'}\) is given on Figure \ref{fig3}.
\begin{figure}[htbp]
 \begin{center}
\includegraphics{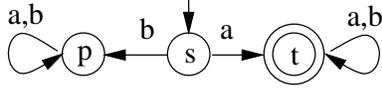}
 \end{center}
\caption{The minimal automaton of \(L'=a^+\{a,b\}^*\).}\label{fig3}
\end{figure}
The seqence \(U\) of Proposition \ref{expexp} can be played by 
\(u_n(t)=2^n\). So here, the Pisot number involved is \(2\) and it is 
multiplicatively independent with \(1+\sqrt{3}\). So from \cite{PB}, 
the only subsets which are simultaneously recognizable in 
\((L,\{a,b,c\},a<b<c)\) and \((L',\{a,b\},a<b)\) are the arithmetic progressions.
}\end{Exa}

\begin{Rem}{\rm Let \(J=a\{a,b\}^*\cup \{a,b\}^*bb\{a,b\}^*\). Notice 
that \(J\) is an exponential language with exponential complement. Its 
minimal automaton \(M_{J}\) is given on Figure \ref{fig4}. 
\begin{figure}[htbp]
 \begin{center}
\includegraphics{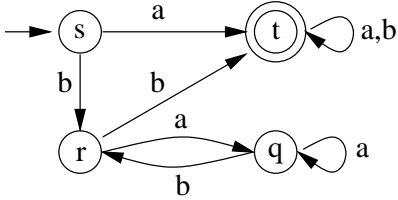}
 \end{center}
\caption{The minimal automaton of 
\(J=a\{a,b\}^*\cup \{a,b\}^*bb\{a,b\}^*\).}\label{fig4}
\end{figure}
We consider the 
numeration system \(S=(J,\{a,b\},a<b)\) and we show that
\begin{itemize}
\item[i)] we cannot find a linear recurrent sequence associated to a 
Pisot number such that the condition (\ref{decomp}) of proposition \ref{expexp} 
is satisfied for all state of \(M_{J}\)
\item[ii)] the set \(X=\{v_n(s) : n \in \N\}\) is \(S\)-recognizable but 
\(2X\) is not.
\end{itemize}
One can check that for all \(n\ge 1\), \(u_n(t)=2^n\) and 
\[u_n(s)=2^n-\frac{\sqrt{5}}{5} \left(\frac{1+\sqrt{5}}{2} \right)^n
+\frac{\sqrt{5}}{5} \left(\frac{1-\sqrt{5}}{2} \right)^n.\]
So i) holds. To check ii), we use the same technique as in Theorem 
\ref{thepol}. One can verify that
\[v_{n+1}(s)-2v_{n}(s)=1-\frac{\sqrt{5}}{5} \left(\frac{1-\sqrt{5}}{2} \right)^n
+\frac{\sqrt{5}}{5} \left(\frac{1+\sqrt{5}}{2} \right)^n\]
has an exponential dominant term. Furthermore, for all \(n\) 
large enough 
there exists \(i\) such that
\[\rho_{J_{b^{i+1}}}(n)=2^{n-i-1}<v_{n+1}(s)-2v_{n}(s)\le 
2^{n-i}=\rho_{J_{b^i}}(n)\]
and \(n-i \to +\infty\) if \(n\to +\infty\). One can conclude as in 
Theorem \ref{thepol}; \(\r{S}(2\, v_n(s))=b^iaz\) with \(\vert 
z\vert=n-i-1\).
}\end{Rem}    

\section{Acknowledgments}
The author would like to warmly thank P.~Mathonet for fruitful discussions 
in the polynomial case 
and also P.~Lecomte for his support and improvements in many 
proofs.
 
\end{document}